\documentclass{template}
\usepackage{url}
\usepackage{xcolor}
\usepackage[skins,theorems]{tcolorbox}
\usepackage{mwe}
\usepackage{csquotes}
\usepackage{amssymb}
\usepackage{graphicx}
\usepackage{array}
\usepackage{colortbl}
\usepackage{xcolor}

\begin{document}

\pagestyle{fancy}

\title{NeoHebbian Synapses to Accelerate Online Training of Neuromorphic Hardware}
\maketitle

\author{S. Pande$^{\ast,1}$},
\author{S.S. Bezugam$^{\circ,2}$},
\author{T. Bhattacharya$^{\diamond,2}$},
\author{E. Wlazlak$^{2}$},
\author{A. Chakaravorty$^{1}$},
\author{B. Chakrabarti$^{\dagger,1}$},
\author{D. Strukov$^{2}$}

\begin{affiliations}
$^{1}$Indian Institute of Technology Madras, Chennai, Tamil Nadu 600036, India\\
$^{\ast}$ee18d704@smail.iitm.ac.in, $^{\dagger}
$bchakrabarti@ee.iitm.ac.in\\

\vspace{0.2cm}
$^{2}$UC Santa Barbara, Santa Barbara, CA 93106-9560, U.S.A\\
$^{\circ}$sbezugam@umail.ucsb.edu, $^{\diamond}$tinish@umail.ucsb.edu\\

\end{affiliations}


\keywords{BPTT, e-prop, NeoHebbian synapses, Online learning, Reinforcement learning, Recurrent spiking neural network, ReRAM, Thermal synapses, Three-factor learning rule}

\begin{abstract}
\noindent
\justifying
Neuromorphic systems that employ advanced synaptic learning rules, such as the three-factor learning rule, require synaptic devices of increased complexity. Herein, a novel neoHebbian artificial synapse utilizing ReRAM devices has been proposed and experimentally validated to meet this demand. This synapse features two distinct state variables: a neuron coupling weight and an \enquote{eligibility trace} that dictates synaptic weight updates. The coupling weight is encoded in the ReRAM conductance, while the \enquote{eligibility trace} is encoded in the local temperature of the ReRAM and is modulated by applying voltage pulses to a physically co-located resistive heating element.
The utility of the proposed synapse has been investigated using two representative tasks: first, temporal signal classification using Recurrent Spiking Neural Networks (RSNNs) employing the e-prop algorithm, and second, Reinforcement Learning (RL) for path planning tasks in feedforward networks using a modified version of the same learning rule. System-level simulations, accounting for various device and system-level non-idealities, confirm that these synapses offer a robust solution for the fast, compact, and energy-efficient implementation of advanced learning rules in neuromorphic hardware.
\end{abstract}
\section{Introduction}
\label{intro}
\justifying

\indent Emulating the biophysical dynamics of the brain by manipulating naturally available physical dynamics lies at the core of neuromorphic computing and is what holds the key to achieving at-par energy efficiency and cognitive capabilities with the human brain \cite{CMead_NC, Mead2020AuthorCH,9395703,schuman2022opportunities,markovic2020physics}. Realizing the full potential of neuromorphic computing requires the development of a computational paradigm that reasonably mimics the structure and functionality of the brain at various levels of abstraction while also being conducive to efficient hardware implementation using state-of-the-art technologies \cite{mehonic_memristorsmemory_2020,roy2019towards,burr2017neuromorphic,upadhyay2019emerging}.The latter can be achieved using memristive devices, which are known for emulating the synaptic functionality due to their ability to tune the conductance to an arbitrary value within its physical dynamic range \cite{Kim20214KmemristorAP,yang_memristive_2013,ielmini_-memory_2018,song_recent_2023}. Additionally, when arranged in crossbar arrays, these devices enable area and energy-efficient in-memory computing by offering massive parallelism. Several studies have reported chip-level demonstrations of neural network accelerators using various memristive devices \cite{LeGallo2022A6M,ambrogio2023analog,sebastian_analog_2023,huang2024memristor}. The former is achieved by adopting spiking neural networks (SNNs). SNNs are known for offering the brain-inspired computational paradigm that comprises approximated neuro-inspired neuron models as activation functions interconnected with synaptic weights and transmit information using asynchronous spike-based events \cite{NIPS2003_rsnn, Li_2023_BIC_survey, Mehonic_A_2022_BIC_review, Ganguly2024}. Thus, using memristor-based hardware to implement SNNs is a compelling alternative for attaining energy efficiency and cognitive performance comparable to those of a biological brain.\par 

SNNs can be trained using the Hebbian learning rules, such as spike-timing-dependent plasticity (STDP) or its variants \cite{Caporale2008SpikeTP, Bi1998SynapticMI, Prezioso2018SpiketimingdependentPL}. In STDP-based learning, the timing and sequence of pre- and post-synaptic spikes determine the magnitude and direction of weight changes \cite{serrano-gotarredona_stdp_2013}. Several experimental demonstrations have shown that SNNs trained with the STDP algorithm can learn to detect temporal correlations within spike trains in an unsupervised manner \cite{milo_resistive_2018,gupta_-chip_2023,prezioso_spike-timing-dependent_2018}. Additionally, large-scale experimental demonstrations have investigated the potential benefits of SNNs \cite{debole2019truenorth, orchard2021efficient, 6805187, gonzalez2024spinnaker2,8998338,qiao2015reconfigurable}. Despite their biological plausibility, SNNs trained using STDP perform poorly on relatively complex tasks primarily due to their focus on local optimization and lack of a global error signal, as seen in artificial neural networks (ANNs) trained with backpropagation \cite{Gerstner2018EligibilityTA}. As a result, the performance of spike-based learning algorithms has often been overshadowed by gradient-based methods used in non-spiking networks. Another significant limitation of SNNs trained with Hebbian learning rules is their inability to model tasks involving long-term temporal dependencies \cite{Bellec2019AST}. While recurrent spiking neural networks (RSNNs) offer a potential solution for modeling such tasks, their training algorithms struggle to assign importance to past neural states for errors observed in the present, making it difficult to determine the necessary adjustments to the network's learnable parameters to achieve desired performance \cite{Bellec2019AST, Bellec2018LongSM}. This issue, known as the temporal credit assignment problem, is not unique to RSNNs but also exists in ANNs. ANNs address this problem using the backpropagation through time (BPTT) algorithm \cite{Werbos_1990_BPTT}. However, BPTT requires unfolding the network and propagating errors backward through time \cite{Werbos_1990_BPTT, LILLICRAP2019_BPTT}, which, while effective for modeling long-term temporal sequences, demands extensive memory, high training time, and significant computational resources, thereby limiting its use in neuromorphic hardware \cite{marschall2020unified}.\par

\indent The eligibility propagation (e-prop) algorithm effectively addresses the temporal credit assignment problem in a biologically plausible way \cite{Bellec2019AST}. Studies have shown that RSNNs trained with e-prop algorithms can learn online and handle complex tasks efficiently \cite{Bellec2019AST}. The e-prop algorithm is a special case of the three-factor learning rule, where synaptic plasticity is influenced not only by the pre-synaptic and post-synaptic neuron signals (as in standard Hebbian learning) but also by an additional third signal. Typically, a three-factor learning rule for synaptic plasticity can be expressed as \cite{Gerstner2018EligibilityTA}: 
\begin{equation*} 
\dot{w} = F(M, pre, post). 
\end{equation*}
In this equation, $\dot{w}$ represents the rate of change of the synaptic weight. The variable $M$ denotes the third signal, and the function $F$ defines the specific learning rule. Three-factor learning rules, including their variants, tackle the issues associated with SNNs by introducing local eligibility traces. These traces, combined with the coupling weight, maintain a fading memory of pre-synaptic activity. Additionally, they make the global error signal (the third signal) locally accessible at the synapse, along with the pre-and post-synaptic signals, facilitating local learning. In the context of temporal modeling tasks, these characteristics eliminate the need for the network to unfold and propagate backward in time, resulting in substantial savings in computational resources and accelerating the training process of neuromorphic hardware.\par
\indent This work focuses on developing a synaptic element tailored for hardware implementation of the e-prop learning algorithm. Our key contributions are as follows: (1) We propose a novel artificial synapse with a two-terminal heater 3D-integrated with a ReRAM cell. This design utilizes intentionally introduced self-thermal coupling between the heater and ReRAM to encode the eligibility trace through the local temperature of the ReRAM, while the non-volatile conductance levels represent the synaptic weights. (2) We provide a comprehensive analysis of the proposed synapse's operation, including its physical mechanisms and various non-idealities. The core operating principle is experimentally validated, and its implementation at the array level is studied within the context of the e-prop algorithm. (3) We present a numerical model to further investigate the synapse's operation and assess its scalability. (4) We evaluate the synapse's performance on two representative tasks using hardware-aware network simulations, accounting for device- and array-level non-idealities. \par
The remainder of the manuscript is organized in the following order: Section \ref{eligibility_based_learning} presents a high-level description of eligibility-based learning, followed by a discussion on the proposed synapse operation, related experimental results, unit cell design, and its array-level operation in Section \ref{thermal_neoHebbian_synapse}. Section \ref{numerical_modeling} covers the numerical modeling of the proposed synapse. Section \ref{benchmark simulations} details system-level simulations used to benchmark the performance benefits of the proposed synapse on two representative tasks: reinforcement learning in SNNs and the more complex TIMIT phoneme classification task in RSNNs. Finally, we conclude by summarizing the scope and limitations of the proposed synapse in Section \ref{discussion}.

\section{Eligibiility-based Learning}
\label{eligibility_based_learning}
\begin{figure}[t!]
\centering
\includegraphics[trim={0.2cm 0.7cm 0.0cm 0.1cm}, clip ,width=0.85\textwidth]{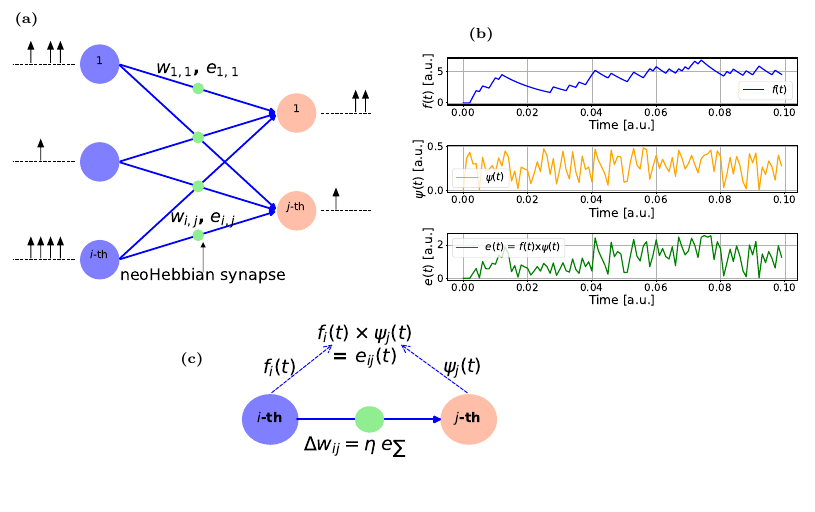} 
\caption{(a) Schematic of a spiking neural network incorporating neoHebbian synapses. (b) The evolution of signals $f(t)$, $\psi(t)$, and $e(t)$ during the dataframe presentation. $f_{i}(t)$ and $\psi_{j}(t)$ represent signals from the $i-$th pre-synaptic neuron and $j-$th post-synaptic neuron, respectively. $e_{ij}(t)$ is obtained by multiplying $f_{i}(t)$ and $\psi_{j}(t)$. (c) Characteristics features of a neoHebbian synapse - computing $e_{ij}(t)$ and accumulating it (i.e., $e_{\sum}$) during the data frame presentation. During the weight update, the weight change ($\Delta w_{ij}$) is proportional to the accumulated $e(t)$. $\eta$ represents the learning rate.}
\label{synapse}
\end{figure}
A high-level description of eligibility-based learning in SNNs utilizing neoHebbian synapses is discussed in this section. Fig.\ref{synapse}(a) shows a schematic of the SNN where input neurons are connected to output neurons using neoHebbian synapses. NeoHebbian synapses exhibit both short-term dynamics and long-term plasticity, characterized by the synaptic \enquote{eligibility trace}(\(e\)) and coupling weight (\(w\)), respectively. The neuronal firing activity at the pre-and post-synaptic neurons dictates the updates in eligibility trace values. These traces serve as temporal markers that record the past activities of the synapse. When the synaptic weights are to be updated, eligibility traces interact with neuromodulator signals to determine the extent and direction (increase or decrease) of synaptic weight changes. In other words, the eligibility trace serves as an additional gating signal that, in conjunction with pre- and post-synaptic activities, influences long-term plasticity and is regulated by the common (two-factor) Hebbian rule.\par 
\indent The computation of the eligibility state ($f(t)$) occurs at the pre-synaptic neuron, while the pseudo-gradient ($\psi(t)$) is computed at the post-synaptic neuron. The product of $f(t)$ and $\psi(t)$ yields the eligibility trace ($e(t)$). The training process operates in batches, where data within each batch, termed as a dataframe, is sequentially processed over $U$ time steps. During the training process, the $e(t)$ is computed and accumulated over the presentation of the dataframe. Fig.\ref{synapse}(b) shows the evolution of signals $f(t)$, $\psi(t)$, and $e(t)$ during the dataframe presentation.\par 
Subsequently, at the end of the dataframe presentation, the coupling weights ($w_{ij}$) are updated proportionally to $e_{\sum}$, where $e_{\sum}$ denote accumulated $e(t)$ is over the dataframe presentation. Overall, the eligibility-based learning approach allows the network to associate specific spike timings with subsequent rewards or punishments, enhancing its ability to perform tasks that require temporal linking of events, such as sequence learning and reinforcement learning, where outcomes are delayed from actions. Detailed equations related to eligibility-based learning are provided in the Appendix section \ref{e_prop_eq} and in Section \ref{benchmark simulations}.\\ 
\indent This work focuses on developing a synaptic device that exhibits both short-term dynamics and long-term plasticity. Additionally, the synapse should be capable of computing and storing \( e(t) \) at the synapse and updating its conductance in proportion to the accumulated \(e(t)\). Fig.\ref{synapse}(c) illustrates the key features of a neoHebbian synapse. A summary of the prior works related to the development of neoHebbian synapses is provided in Appendix Section \ref{prior_works}.

\begin{figure}[h!]
\centering
\includegraphics[trim={0.4cm 0.8cm 0.5cm 0.31cm}, clip ,width=0.81\textwidth]{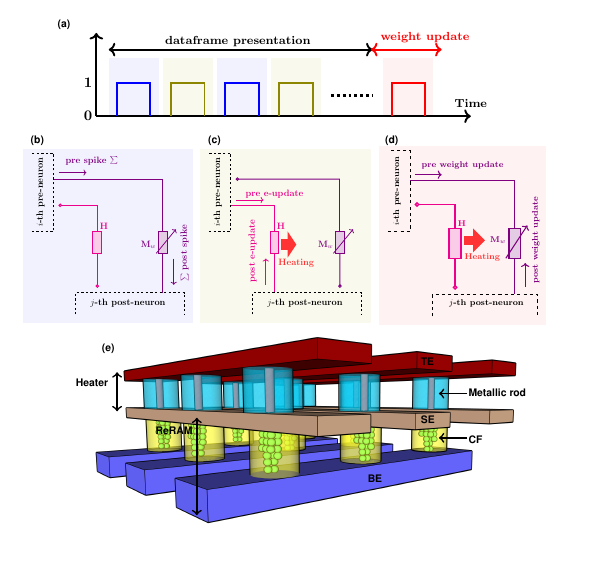}
\caption{High-level description of the thermal neoHebbian synapse operation: (a) Three key stages involved in the training operation of e-prop: spike integration (shaded blue), eligibility-update (shaded olive), and weight update (shaded red). Arrangement of the heater and ReRAM during (b) Spike integration, (c) e-update, and (d) Weight update phases. The red arrow depicts the thermal coupling between the heater and ReRAM. (e) Design of a crossbar array illustrating 3D-integrated heater and ReRAM cells.}
\label{main_idea}
\end{figure}
\section{Thermal NeoHebbian Synapse}
\label{thermal_neoHebbian_synapse}
The high-level functionality of the proposed synapse in the context of the e-prop training algorithm is outlined in the following. The training process in e-prop consists of three key stages: the spike integration (or inference) phase, the eligibility update (e-update) phase, and the weight update phase. The spike integration and e-update occur during the data frame presentation, while the weight update is executed after the data frame presentation (refer to Fig.\ref{main_idea}(a)). Consider the heater ($H$) and ReRAM ($M_{w}$) arrangement as shown in Fig.\ref{main_idea}(b). In this, $M_{w}$ acts as coupling weight and transmits weighted current spikes from $i$-th pre-neuron to $j$-th post-neuron during the spike integration phase, as shown in Fig.\ref{main_idea}(b). During the e-update phase, the heater receives appropriate signals such as $f(t)$ and $\psi(t)$ from pre- and post-neurons, respectively, resulting in the rise in heater temperature due to Joule heating. Due to the thermal coupling between the heater and ReRAM, the local temperature of ReRAM increases proportionally to the dissipated power. The e-update phase is depicted in Fig.\ref{main_idea}(c). These operations are repeated at every step during the dataframe presentation in the training process. Finally, at the end of the data frame presentation, the accumulated temperature rise in ReRAM represents \enquote{$e_{\sum}$}, thus satisfying the requirement of computing and storing the eligibility trace at the synapse. Subsequently, during the weight update, a fixed-amplitude programming pulse is applied, which induces a conductance change ($\Delta G_{w}$) proportional to the accumulated temperature rise ($e_{\sum}$). Essentially, the temperature-dependent switching behavior of the ReRAM is exploited to update the weights proportional to $e_{\sum}$.\par
\indent
To implement the characteristic features highlighted by the high-level functionality of the thermal neoHebbian synapse, we propose the integration of a two-terminal heater cell with the ReRAM device, as depicted in Fig.\ref{main_idea}(e). The heater element comprises an insulating layer sandwiched between two metallic layers: the top electrode (TE) and the shared electrode (SE). A metallic nanorod connects the TE and SE. 
Upon applying a voltage between the TE and SE, a substantial current flows through the metallic nanorod, which has a high electrical conductivity, resulting in localized Joule heating. Due to the high thermal conductivity of the SE, strong thermal coupling is established between the heater element and the ReRAM. The ReRAM switching layer is sandwiched between the SE and the bottom electrode (BE). To mitigate lateral heat diffusion to adjacent cells, the nanorod structure is surrounded by an electrically and thermally insulating layer. The desired properties of the heater are akin to the heater used in mushroom-type phase-change memory technologies \cite{pcm_1,pcm_2,pcm_3}. Consequently, suitable materials for the electrode layers include W, TiN, and TaN, and for the insulating layers, materials such as SiO$_{2}$, HfO$_{2}$, and TiO$_{2}$. This design minimizes area footprint by 3D-integration of the heater and the ReRAM cells. Details about the fabricated ReRAM layer stack and its electrical characteristics are presented in the following section. \par

\subsection{Experimental Results}
\label{experimental_results}
\begin{figure}[h!]
\centering
\includegraphics[trim={2.2cm 1.8cm 0.3cm 0.5cm}, clip ,width=1.1\textwidth]{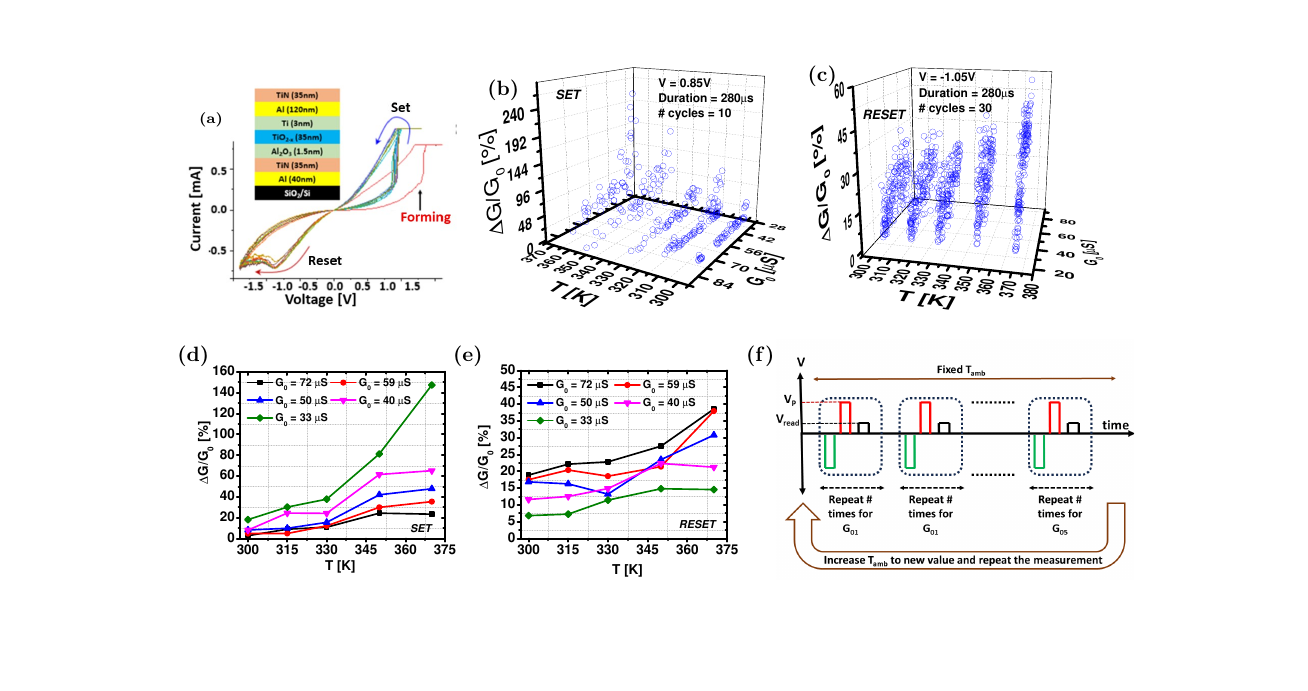} 
\caption{(a) Representative I-V curves measured with quasi-static DC voltage sweep at 1V/s on 250$\times$250nm$^2$ area devices. The inset provides the device stack details. Normalized percentage conductance change as a function of the initial conductance and ambient temperature is shown for the (b) SET and (c) RESET processes. The average normalized percentage conductance change as a function of ambient temperature for the SET and RESET processes is presented in (d) and (e), respectively. (f) Illustration of the measurement protocol used to obtain the data is shown in (b) and (c). The green pulse depicts the multiple SET, RESET, and read pulses required to reprogram the device to the same $G_\mathrm{0}$. $V_\mathrm{P}$ and $V_\mathrm{read}$, respectively, denotes fixed programming pulse used to measure $\Delta G$ and read pulse.} 
\label{measured_data}
\end{figure}
\noindent Metal oxide memristors were fabricated using a similar process to our previous work \cite{Kim20214KmemristorAP}, which involved etch-down processes, and UV lithography for patterning, DC-mode magnetron sputtering for electrode deposition, and thermal annealing in forming gas to adjust non-stoichiometry. However, in this work, an oxide bilayer stack was formed using ALD to simplify the fabrication process and mitigate issues related to thickness and composition variations in the sputtering targets. Fig.\ref{measured_data}(a) shows typical $I$-$V$ switching curves from these devices, with the inset providing details of the ReRAM layer stack. These devices exhibit low forming voltages ($\sim$2V), switching voltages ($\sim$1V), and an on/off ratio of $\sim$20 at a read voltage of 0.1V.\par
We now examine the temperature-dependent switching characteristics within the context of thermal neo-Hebbian synapse operation. In the proposed design, the heater and ReRAM cells are 3D-integrated, with the structure optimized to maximize thermal coupling between them. The close proximity of the heater and ReRAM, along with the high thermal conductivity of SE, allows for simplification of experimental measurements by emulating the heater's role through modulation of the ambient temperature. Our investigation focuses on understanding the normalized conductance change $(\Delta G/G_{0})$ induced by a fixed voltage pulse as a function of the initial programmed conductance ($G_{0}$) and ambient temperature ($T$). Fig.\ref{measured_data}(b) and Fig.\ref{measured_data}(c) present $(\Delta G/G_{0})$ as a function of $G_{0}$ and $T$ for the SET and RESET processes, respectively.\par
\indent The measurement protocol employed during the experiments is as follows: First, the ReRAM is programmed to a target initial conductance $G_{0}$ with 5\% tuning accuracy, using the tuning algorithm described in \cite{alibart2012high}. Next, a programming pulse of fixed amplitude and duration is applied, followed by a read pulse to measure the conductance change ($\Delta G$). The device is then reprogrammed to the same $G_{0}$, and the measurement is repeated multiple times to collect several data points for each $G_{0}$. This procedure is repeated for all specified $G_{0}$ values. Afterward, the ambient temperature increases and the entire process is repeated. Fig.\ref{measured_data}(f) illustrates the measurement protocol.
For the SET process, data were obtained using a pulse with an amplitude of 0.85V and a duration of 280$\mu$s, with measurements repeated 10 times for each combination of $G_{0}$ and $T$, yielding a total of 250 data points. For the RESET process, a pulse of -1.05V with a duration of 280$\mu$s was applied, and measurements were repeated 30 times for each combination of $G_{0}$ and ambient temperature, resulting in a total of 750 data points. Fig.\ref{measured_data}(d) shows the average normalized conductance change for the SET process, where each data point represents the average of 10 measurements, while Fig.\ref{measured_data}(e) shows the corresponding average values for the RESET process, with each data point representing the average of 30 measurements.\par
\indent It was observed that during the SET process, the conductance change reaches its maximum when \( G_{0} \) is close to the device’s lowest conductance state and its minimum when \( G_{0} \) approaches the highest conductance state. In contrast, during the RESET process, the conductance change is at its maximum when \(G_{0}\) is near the highest conductance state and at its minimum when \(G_{0}\) is close to the lowest conductance state. This behavior is attributed to the fixed dynamic conductance range in ReRAM devices, leading to saturation of conductance updates as the conductance approaches the extremes of its dynamic range.

\subsection{NeoHebbian Synapses: Unit Cell and Array level Operation}
\label{neoHebbian_synapse}

\begin{figure}[t!]
\centering
\includegraphics[trim={0.35cm 0.35cm 0.35cm 0.32cm}, clip ,width=0.9\linewidth]{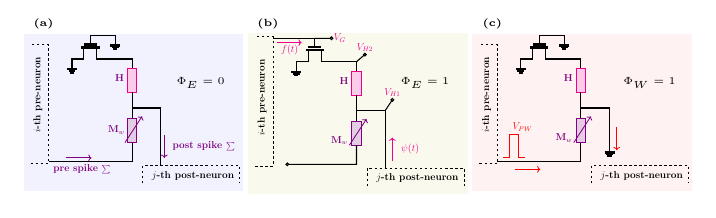}    
\caption{1$T$-1$H$-1$M$ unit cell implementation of the thermal neoHebbian synapse. During the dataframe presentation, the operation of the synapse is time multiplexed between (a) Spike integration - $\phi_E$ = 0 and (b) e-update - $\phi_E$ = 1 phase (c) Weight update is performed at the end of the dataframe - $\phi_W$ = 1. The appropriate biasing conditions for each phase are shown in the schematic.}
\label{1T_1H_1M_working}
\end{figure}
\indent The unit-cell implementation of thermal neo-Hebbian synapses is shown in Fig.\ref{1T_1H_1M_working}. This unit cell consists of one transistor, one heater, and one ReRAM device, hence referred to as the 1$T$-1$H$-1$M$ configuration. During the spike integration phase (\(\phi_E = 0\)), the memristor serves as a coupling weight, facilitating the transmission of weighted current spikes from the pre-neuron to the post-neuron, while the transistor remains off, as depicted in Fig.\ref{1T_1H_1M_working}(a).\par
\indent In the e-update phase (\(\phi_E = 1\)), the eligibility state \(f(t)\) and the pseudo-gradient \(\psi(t)\), calculated at the pre-and post-synaptic neurons, respectively, are applied across the heater. Simultaneously, the memristor (\(M_w\)) is decoupled from the pre-neuron, as illustrated in Fig.\ref{1T_1H_1M_working}(b). The applied voltage signals result in Joule heating, raising the temperature of the heater, which subsequently increases the local temperature of the ReRAM through thermal coupling. The total eligibility contribution to the weight update is stored as the cumulative temperature rise in the ReRAM during data presentation. In the final weight update phase (\(\phi_W = 1\)), a fixed-amplitude programming pulse is applied, leading to a change in conductance \(\Delta G\), which is proportional to the accumulated eligibility trace (\(e_{\sum}\)), as shown in Fig.\ref{1T_1H_1M_working}(c).  Shared usage of one of the electrode terminals between the heater and the memristor in a 1$T$-1$H$-1$M$ configuration requires time-division multiplexing between the $\phi_E$ = 0 and $\phi_E$ = 1 phases. However, introducing an additional transistor in the synaptic cell can eliminate the need for time-division multiplexing between the spike integration and the e-update phases in the 1$T$-1$H$-1$M$ design, albeit with the trade-off of reduced density \cite{IEDM_shu2023}. \par
\indent An essential feature of the noeHebbian synapse is the local computation and storage of $e(t)$. As illustrated in Fig.\ref{synapse}(c), during e-prop operation, $e(t)$ is computed as the product of $f(t)$ and $\psi(t)$. In the context of the thermal neoHebbian synapse, this implies that the local temperature of $M_{w}$ should increase proportionally to the product of voltage signals $f(t)$ and $\psi(t)$. The 1$T$-1$H$-1$M$ unit cell allows local computation and storage of $e(t)$ through appropriate biasing and voltage scaling.\par
\noindent During $\phi_E$ = 1, assuming the transistor is operating in the triode regime, the drain current ($I_\mathrm{D}$) is given by, 
\begin{equation*}
    I_\mathrm{D}(t)=k\left(V_\mathrm{GS}(t)-V_\mathrm{TH}\right) V_\mathrm{H2}(t).
\end{equation*}
The voltage drop across the heater can be expressed in terms of the drain current as: 
\begin{equation*}
    V_\mathrm{H1}(t) - V_\mathrm{H2}(t) =  I_\mathrm{D}(t) R. 
\end{equation*}
Here, $R$ denotes heater electrical resistance, and $k$ is a transistor-related parameter. Consequently, the power dissipated across the heater is expressed as:
\begin{equation}
P_\mathrm{H}=\frac{1}{R} V_\mathrm{H1}^2 \frac{\left(Rk V_{\mathrm{OV}}\right)^2}{\left(1+Rk V_{\mathrm{OV}}\right)^2} \:.
\end{equation}
Now, $V_\mathrm{H1}$ and $V_{\mathrm{OV}}$ are scaled as follows, 
\begin{equation}
 V_\mathrm{H1} \propto \sqrt{\psi(t)} \quad \& \quad V_\mathrm{OV} \propto \frac{\sqrt{f(t)}}{{Rk}(1-\sqrt{f(t)})}.
\end{equation}
$V_{\mathrm{OV}}$ denotes transistor overdrive voltage. This voltage scaling performed at the neuron site ensures that the dissipated power across the heater and, consequently, the temperature rise at the heater follows the desired proportionality:
\begin{equation}
P_\mathrm{H} \propto \psi(t) \times f(t) \quad \text{;} \quad \therefore \; \Delta T \propto \psi(t) \times f(t).
\end{equation}
\noindent Due to thermal coupling, the power dissipated across $M_\mathrm{w}$ is proportional to $P_\mathrm{H}$; consequently, the local temperature of $M_\mathrm{w}$ increases in relation to the product of $\psi(t)$ and $f(t)$. The more details on the related equations are provided in Appendix Section \ref{elig_comp}.\par

\begin{figure}[t!]
\centering
\includegraphics[trim={0.35cm 0.32cm 0.35cm 0.32cm}, clip ,width=0.87\textwidth]{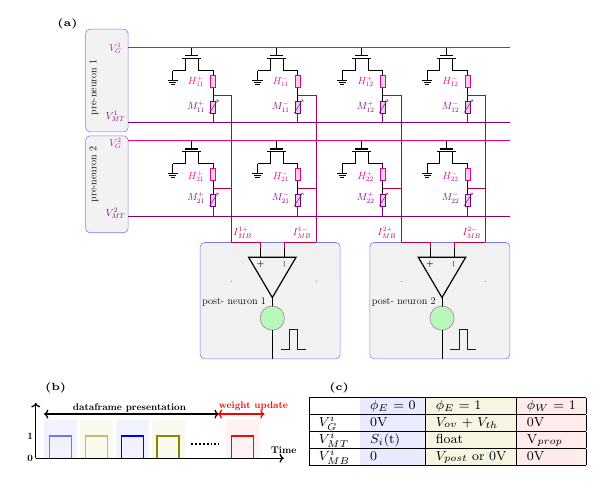}
\caption{(a) Differential mode crossbar-array implementation of the 1$T$-1$H$-1$M$ design. (b) Key stages in the operation of e-prop: spike integration ($\phi_E$ = 0), e-update ($\phi_E$ = 1), and weight update ($\phi_W$ = 0). (c) Biasing condition at the respective terminal during these phases. }
\label{crossbar_implementation}
\end{figure}

\indent We now discuss the array-level operation of the proposed synapse. The array-level implementation of the 1$T$-1$H$-1$M$ design involves synaptic cells arranged in a differential configuration, as shown in Fig.\ref{crossbar_implementation}(a). In this setup, two sets of synapses are utilized. The net synaptic conductance is given by $G = G^+ - G^-$, where $G^+$ denotes the total conductance of memristor $M^{+}_{w}$, and $G^-$ represents the total conductance of memristor $M^{-}_{w}$. The net conductance $G$ can be increased (decreased) by potentiating (depressing) $G^+$ or depressing (potentiating) $G^-$ \cite{Boybat2017NeuromorphicCW, Differential_Reram}.\par

Fig.\ref{crossbar_implementation}(c) shows the voltage bias applied at respective terminals during spike integration, e-update, and weight update. The e-update and subsequent weight update operation in the differential mode operate as follows: During e-prop operation, $f(t)$ maintains a strictly positive value, while $\psi(t)$ can be either positive or negative. Depending on the sign of $\psi(t)$, the e-update operation is directed towards either heater $H^{+}$ or heater $H^{-}$. For instance, when $\psi(t) > 0$, $f(t)$ and $\psi(t)$ are applied to the terminals of heater $H^{+}$, as illustrated in Fig.\ref{1T_1H_1M_working}(b). Conversely, if $\psi(t) < 0$, the update is directed towards heater $H^{-}$. During the $\phi_W$ = 1 phase, a fixed amplitude programming pulse is simultaneously applied to both memristors $G^{+}$ and $G^{-}$. Consequently, the resulting change in conductances, denoted as $\Delta G^{+}$ and $\Delta G^{-}$, is directly proportional to the local temperature increase at the respective synapse during the e-update phase. Therefore, the net change in conductance ($\Delta G$) is calculated as $\Delta G = \Delta G^{+} - \Delta G^{-}$. It's important to note that the voltage pulse used during $\phi_W$ = 1 induces insignificant conductance change if the local temperature rise at the memristor is negligible. The array-level implementation depicted in Fig.\ref{crossbar_implementation}(a) facilitates parallelism, leading to substantial time savings during training. For example, during $\phi_E$ = 1, eligibility is updated concurrently for all the elements in the array, followed by concurrent weight update at the end of the dataframe in $\phi_W$ = 1.

\vspace{-0.2cm}
\section{Numerical Modeling}
\label{numerical_modeling}
\begin{figure}[t!]
\centering
\includegraphics[trim=0.2cm 0.2cm 0.2cm 0.2cm, clip=true,width=0.9\linewidth]{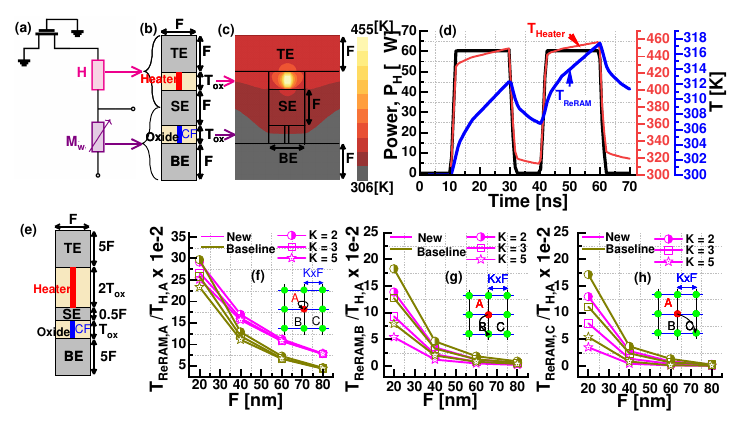}  
\caption{(a) 1T-1H-1M synapse unit cell. (b) Modeled geometry for electrothermal analysis. (c) Temperature contours calculated at $t$ = 60ns for $F$ = 60nm and overlaid on the modeled geometry, showing the thermal self-thermal coupling between the heater and ReRAM. (d) Transient temperature evolution depicting $\phi_{E}$ = 1 phase. (e) Modified geometry that improves self-thermal coupling and reduces the thermal crosstalk. A comparison of thermal coupling coefficients for the structure shown in (b) \& (e) is shown in (f,g,h). (f) self-thermal coupling between heater and ReRAM at location \textit{A}. Thermal crosstalk between the heater at location \textit{A} and ReRAM at location \textit{B} and \textit{C} is shown in (g) \& (h), respectively. The inset shows the schematic of the modeled 3$\times$3 crossbar array, where $K$ denotes crossbar pitch.}
\label{numerical_sim}
\end{figure}

The electrothermal effects are critical in the operation of the proposed synapse and are further investigated using the numerical model. Fig.\ref{numerical_sim}(a) shows the schematic of the 1$T$-1$H$-1$M$ synapse, and the corresponding modeled geometry considered for electrothermal simulation is shown in Fig.\ref{numerical_sim}(b). The oxide thickness ($T_{\mathrm{ox}}$) is assumed to be 30nm for both the heater and ReRAM (see Fig.\ref{numerical_sim}(b)). All other dimensions are marked in minimum feature size ($F$). The time-dependent temperature profile within the device is obtained by solving the transient heat flow equation. More details on the numerical model are provided in the Appendix Section \ref{num_model}.\\
Fig.\ref{numerical_sim}(d) illustrates the e-update phase, showing the transient temperature evolution at the heater and ReRAM in response to the dissipated power at the heater. Temperature contours calculated at \( t = 60 \)ns are overlaid on the modeled geometry and shown in Fig.\ref{numerical_sim}(c), highlighting the thermal coupling between the heater and ReRAM. Overall, Fig.\ref{numerical_sim}(c) and Fig.\ref{numerical_sim}(d) validate the capability to encode the \enquote{eligibility state} in the form of local temperature, as the local temperature of the ReRAM can be modulated by applying heating pulses at the heater.\\
\indent We now examine potential sources of non-idealities that might influence the performance metrics of e-prop. For instance, the accumulated eligibility (\(e_{\sum}\)) is expected to remain constant, even in the absence of activity (see eq.\ref{eq8} in Appendix Section \ref{e_prop_eq}). However, in the proposed synapse, \(e_{\sum}\) decreases due to natural temperature decay in the absence of heating pulses (refer to Fig.\ref{numerical_sim}(d)). Ideally, this suggests that the desired $\tau_\mathrm{TH}$ should tend towards infinity. However, it will be evident in the next section that the desired value of $\tau_\mathrm{TH}$ depends on the target application, and, in fact, this eligibility decay could be useful in certain cases. Additionally, it's important to note that when the pulse width ($t_\mathrm{PW}$) exceeds the device thermal time constant ($\tau_\mathrm{TH}$), the ReRAM temperature reaches the steady state, impeding further e-updates. Thus, the $t_\mathrm{PW}$ should be shorter than $\tau_\mathrm{TH}$ during the $\phi_E = 1$ phase to update the eligibility state continuously.\\
\indent Another important non-ideality is the thermal crosstalk during the $\phi_E=1$ phase. The unintentional rise in the local temperature of the neighboring synapses could result in an erroneous e-update. Therefore, we define the thermal crosstalk coefficient as the ratio of temperature rise at the adjacent synapses to temperature rise at the heater in response to heating pulse in $\phi_E$=1 phase. To mitigate the issue of thermal crosstalk, modifications are made to the device structure, as illustrated in Fig.\ref{numerical_sim}(e). These modifications include reducing the distance between the heater and ReRAM to enhance the desired self-thermal coupling and increasing the thickness of both the top and bottom electrodes to slow the propagation of heat flux toward neighboring devices, thereby reducing unintentional thermal crosstalk. Further, the thermal crosstalk coefficients for the structures shown in Fig.\ref{numerical_sim}(b,e) are compared for different values $F$ and $K$. Fig.\ref{numerical_sim}(f) shows that the new design increases the self-thermal coupling and reduces thermal crosstalk, as shown in Fig.\ref{numerical_sim}(g,h). 
The effects of these non-idealities, including eligibility decay and thermal crosstalk, are examined in detail in the benchmark simulations section.
\section{Benchmark Simulations}
\label{benchmark simulations}
\subsection{Case Study \#1: Reinforcement Learning in SNNs}
\begin{figure}[t!]
\centering
\includegraphics[width=0.8\textwidth]{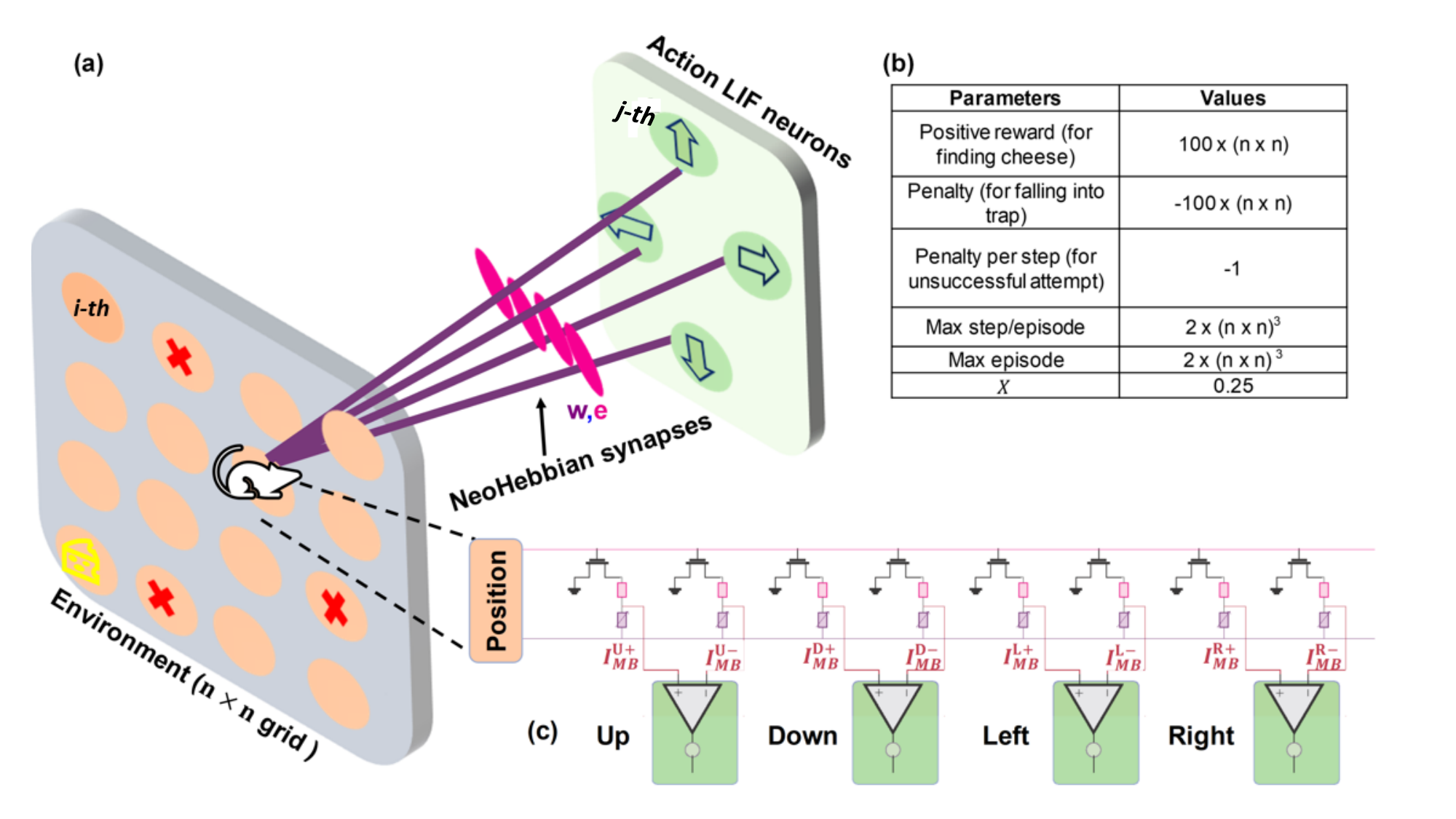}
\caption{(a) Schematic of SNN used for illustrating reinforcement learning using neoHebbian synapse. Here, the agent, depicted as a mouse, must navigate through a $n \times n$ grid to locate cheese and avoid traps. The traps are shown using the red color cross. NeoHebbian synapse connects every $i$-th neuron in a 
$n \times n$ grid to every $j$-th neuron in the output layer (connection for only one $i$-th neuron is shown to avoid clutter). (b) Parameter values used in the simulations. (c) Schematic of array level implementation of the network shown in (a). Here $n \times n$ denotes the grid length.}
\label{TFL}
\end{figure}

This case study discusses the use of neoHebbian synapses in training SNNs for tasks related to reinforcement learning. Specifically, we explore a scenario where a virtual agent resembling a mouse navigates a maze in search of cheese while avoiding traps (see Fig.\ref{TFL}(a)). The maze is structured like a $n \times n$ grid, where the mouse's current position defines its state. At each step, the agent, or mouse, is limited to a single action: moving in one of four directions \enquote{up}, \enquote{down}, \enquote{left}, or \enquote{right}. An episode in this context refers to a single run of the agent through the maze, from start to termination. Each episode begins with the agent randomly placed within the maze and terminates when the agent either finds cheese or encounters a trap. Following each episode, a new round commences from a randomly selected location within the maze. The agent is trained over multiple episodes, learning from past experiences to improve performance. The reward system is designed to maximize the agent's chances of finding cheese, offering positive rewards for success and penalties for falling into traps. Additionally, each action made without finding cheese results in a minor penalty. The respective parameters are summarized in the table shown in Fig.\ref{TFL}(b).\\
\indent In each episode, the agent navigates the grid by making decisions at every timestep. The assumed grid arrangement is akin to an input layer (environment position/state) where each location on the grid is connected to four Leaky-Integrate and Fire (LIF) neurons in the output layer (representing action) using thermal neoHebbian synapse as shown in Fig.\ref{TFL}(a). The LIF neurons drive the agent's decision-making process at each time step in the output layer. For example, suppose the LIF neuron corresponding to the action \enquote{up} direction in the output layer exhibits the highest membrane potential, which is influenced by both the current grid position of the mouse and the accumulated potential from the previous state. In that case, the mouse will move in the \enquote{up} direction. Further, homeostasis is applied on the most recent action (output LIF neuron with highest membrane potential) by decreasing the membrane potential by half.\\
\indent During an episode, the eligibility value is updated at every time step according to the following procedure. Referring to Fig.\ref{TFL}(a), suppose the agent is positioned at the $i$-th neuron in an $n \times n$ grid. If the membrane potential of the $j$-th action neuron is the highest, then the eligibility value for the synapse connecting the $i$-th position neuron with the $j$-th action neuron is increased. Therefore, the updated eligibility takes the form, 
\begin{equation}
e_{i,j}(t) = e_{i,j}(t-1) + 1.    
\end{equation}
\indent And the eligibility values of all other synapses undergo the leakage similar to works \cite{espino2024rapid,galloni2024neuromorphic}, 
\begin{equation}
e_{i,j}(t) = \gamma \; e_{i,j}(t-1).
\end{equation}
\indent Here $\gamma$ is the discount factor, and it ranges between 0 to 1. The synaptic weights are updated at the end of each episode in proportion to the accumulated rewards and eligibility value, as shown in the following equation. 
\begin{equation}
   \Delta W_{ij} = \bigg(\frac{1}{1+e^{-r}}-\chi\bigg)\cdot \sum^{episode} e_{ij}.  
\end{equation}

\indent Here, \textit{r} denotes the ratio of the accumulated rewards and penalties in an episode relative to the highest positive reward. The weight update procedure ensures that both rewards and recent actions are considered during learning.
The neuron membrane potential and eligibility values are reset to zero at the beginning of each episode. We scale the worst and best rewards proportional to grid length to standardize rewards across various grid sizes. This scaling approach ensures consistency in the reward magnitudes relative to the size of the grid (see Fig.\ref{TFL}(b)).\\
\indent As discussed in earlier sections, the natural decay of temperature in the absence of heating pulses represents an important non-ideal aspect. This decay results in a reduction in accumulated eligibility, which is typically expected to remain constant until the weight update occurs (see eq.(\ref{eq8}) in appendix section \ref{e_prop_eq}). However, in the context of the reinforcement learning scenario, this non-ideality facilitates the realization of the discount factor $\gamma$. This factor allows the agent to prioritize recent experiences while gradually diminishing the significance of older ones, which is crucial for effectively adapting to the changing challenges presented by the environment.\\ 
\begin{figure}[t!]
    \centering
    \includegraphics[width=0.9\textwidth]{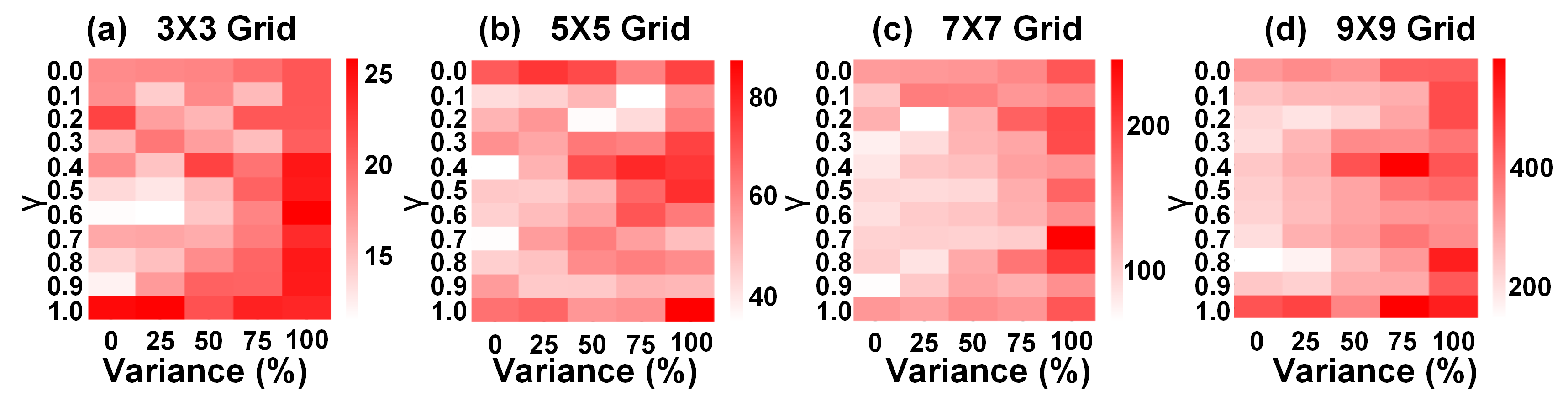}
    \caption{Heatmap compares the average number of episodes required to reach the learning benchmark across different grid sizes: (a) $3 \times 3$ grid, (b) $5 \times 5$ grid, (c) $7 \times 7$ grid, and (d) $10 \times 10$ obtained by considering the impact of temperature decay and memistor variability. Simulation results, which are the mean of 20 runs for each unique combination of $\gamma$ and variability in each grid size, are obtained using a pair of ReRAM devices in a differential configuration to implement a synapse, with each ReRAM device assumed to have 7-bit precision.}
    \label{TFL_1}
\end{figure}

\begin{figure}[h!]
    \centering
    \includegraphics[trim={0.5cm 9cm 1cm 3cm}, clip ,width=1\textwidth]{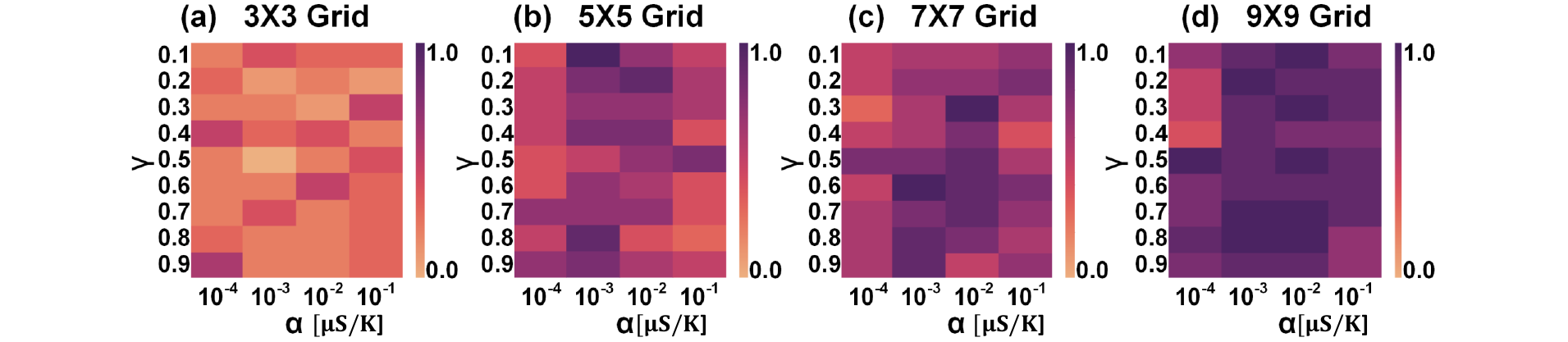}
    \caption{The heatmaps illustrate success ratios for a spiking neural network agent's training across various grid sizes with $\gamma$ (y-axis) controls temperature decay, while $\alpha$ is varied along the x-axis. Lighter shades indicate lower success ratios. Simulation results, which are the mean of 20 runs for each unique combination of $\gamma$ and $\alpha$ in each grid size, are obtained using a pair of ReRAM devices in a differential configuration to implement a synapse, with each ReRAM device assumed to have 7-bit precision.}
    \label{TFL_2}
\end{figure}

\indent Benchmark simulations are performed on various $n\times n$ grids ($n$ = 3,5,7,10) to investigate the influence of temperature decay on the agent's learning ability. In this context, \enquote{learning} refers to the agent's ability to earn five consecutive positive rewards. Fig.\ref{TFL_1} compares the average number of episodes required to reach this learning benchmark across different grid sizes, considering the effects of temperature decay and ReRAM variability. For instance, in scenarios where $\gamma$=0, eligibility accumulation is null due to rapid temperature decay, while $\gamma$=1 signifies no reduction in accumulated eligibility owing to extremely slow temperature decay. When $\gamma$=0, it's expected that there would be an increase in the average number of episodes required to reach the learning benchmark, as the agent doesn't consider prior experiences while making decisions. Interestingly, it's observed that the average number of episodes needed to reach the learning benchmark for $\gamma$=1 is also higher across all grid sizes, indicating that the agent struggles to achieve the benchmark if the temperature decay is extremely slow. This effect is particularly pronounced in larger grid sizes (n=7,10), where complexity and the number of possible paths are higher. The agent learns faster with optimal temperature decay, as indicated by the optimal $\gamma$ value in the heatmap, is considered. Therefore, the seemingly non-ideal effect of temperature decay proves beneficial in reinforcement learning, as it enables the agent to prioritize recent experiences and gradually diminishes the importance of past experiences. Moreover, increased ReRAM variability hampers the agent's learning process, as evidenced by the rise in the average number of episodes needed to reach the learning benchmark with increasing variability. Details about the variability model used in our simulations are provided in the appendix section \ref{benchmark_sim}.\\
\indent Next, with a fixed 50\% variability, we analyze the impact of temperature-induced changes in ReRAM conductance, modeled as \( W = W(1 + \alpha(T - T_\mathrm{amb}))\), where \( W \) is ReRAM conductance \cite{9047174}. A lower \(\alpha\) value is typically preferred in these applications. The heatmap in Fig.\ref{TFL_2} shows the agent's success ratio within a maximum number of episodes for each grid size. The success ratio represents the number of times the agent reached the learning benchmark for each unique combination of \(\gamma\) and \(\alpha\), divided by the maximum number of times the benchmark was reached.  Fig.\ref{TFL_2} shows that the influence of \(\alpha\) becomes less significant with increasing grid sizes, potentially due to the increased redundancy. This observation underscores the potential for efficient operation in dense arrays, a capability that will be further explored in a subsequent case study involving more complex networks.\\

\begin{figure}[t!]
\centering
\includegraphics[trim={0 0 0 0}, clip ,width=0.6\textwidth]{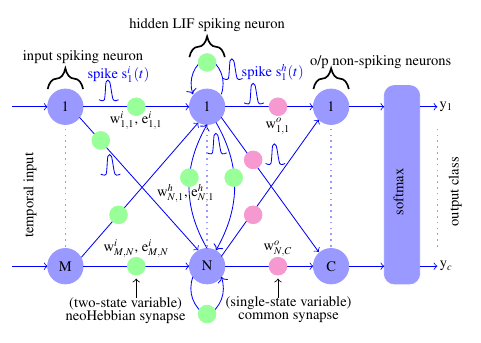}
\caption{The schematic of the fully connected RSNN with one hidden layer. The input and hidden layers consist of spiking LIF neurons. NeoHebbian synapses connect the input layer with the hidden layer and recurrent connection within the hidden layer. The output readout and hidden layers are connected using (common) Hebbian synapses. $w^i_{ij}$, $w^h_{ij}$, $w^o_{ij}$ denote the synaptic weights in the input, hidden and output layer. $e_{ij}$ and $s(t)$ denote the stored eligibility value and spikes from LIF neurons, respectively.}
\label{rsnn_sch}
\end{figure}
\subsection{Case Study \#2: RSNNs for Phenome Classification}
\begin{figure}[t!]
\centering
\includegraphics[width=1\textwidth]{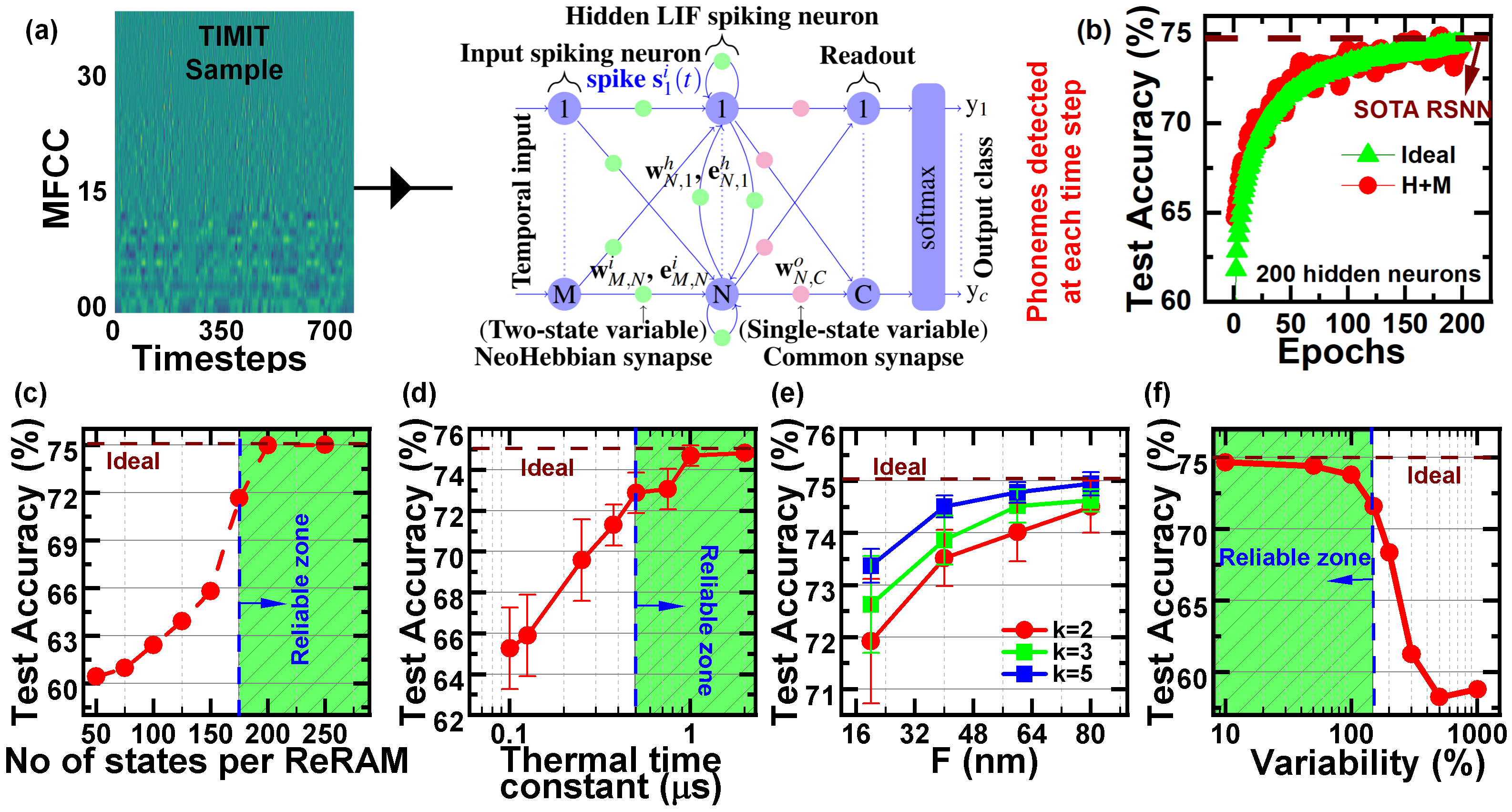}  
\caption{(a) A sample from the TIMIT data set applied to the input layer of the modeled RSNN used for the 
TIMIT phenome detection task. (b) Ideal (software modeled) synapse and proposed synapse test accuracy comparison (c,d,e,f) shows the test accuracy sensitivity towards various sources of non-idealities such as (c) Bit precision, (d) Thermal decay, (e) Thermal crosstalk, (f) Variability.}
\label{benchmark_simulations}
\end{figure}

In this case study, we investigate the performance of the thermal neoHebbian synapse in the TIMIT phoneme classification task. We employ the e-prop algorithm to conduct online training of recurrent spiking neural networks (RSNNs) featuring thermal neoHebbian synapses on the TIMIT dataset. TIMIT phoneme recognition serves as a standard measure for assessing the temporal processing capabilities of recurrent neural networks \cite{garofolo1993timit}. The dataset consists of acoustic speech signals from 630 speakers across eight dialect regions of the USA. The objective is to identify the spoken phoneme among 61 phonemes within each 10ms time frame.\\ 
\indent The schematic representation of the modeled RSNN network used in this study is illustrated in Fig.\ref{benchmark_simulations}(a), comprising 39 input neurons, one hidden layer with 200 LIF neurons, and an output layer consisting of 61 neurons, operating over an average of 700 time steps per sample during inference. The input data is encoded following the procedure outlined in \cite{Bellec2019AST}, and one sample input data is shown in Fig.\ref{benchmark_simulations}(a).
The LIF spiking neurons in the input layers are connected to the hidden layer LIF neurons via neoHebbian synapses. Hidden layer LIF neurons are recurrently connected to themselves and other neurons by neoHebbian synapses. An output (readout) layer of non-spiking neurons is connected to the hidden layer with common (Hebbian) synapses. Spikes coming from the input layer ($s^{i}(t)$) and recurrent connections ($s^{h}(t)$) updates the membrane potential of the hidden layer neurons. The training operation is performed as follows. During the e-update phase, the eligibility contribution is computed at each time step and accumulated at synapse during the presentation of $U$-step long input dataframe as follows, 
\begin{equation}
e_{\Sigma} = \sum_{t=1}^U e_{ij}(t), \text{ where } e_{ij}(t) = f_{i}(t) \times \psi_{j}(t).
\end{equation}
\indent Here, $f(t)$ and $\psi(t)$ signals are provided from pre-synaptic and post-synaptic neurons, respectively. The network loss is calculated at the non-spiking output neurons, and batch-mode stochastic gradient descent is used to update the output layer weights ($w^o$). The neoHebbian synapse ($w^{i / h}$) are updated according to  
\begin{equation}
\Delta w^{i / h} = \eta e_{\Sigma}.
\end{equation}
Here, parameter $\eta $ denotes the learning rate. The values of $e_{\Sigma}$, $f(t)$, and $\psi(t)$ are set to zero at $t=0$, i.e., before the presentation of a new training dataframe. Note that updates to the readout weights ($w^o$) and input/recurrent neoHebbian weights ($w^{i / h}$) occur exclusively at the end of each dataframe. Appendix Section \ref{e_prop_eq} provides more details on the key equations used in this case study.\\

Fig.\ref{benchmark_simulations}(b) compares the training accuracy obtained using ideal (software-modeled) synapses and the proposed neoHebbian synapses. The proposed synapses perform comparably to ideal synapses, assuming floating-point precision. We then investigated the dependence of test accuracy on synapse bit precision, as shown in Fig.\ref{benchmark_simulations}(c). Our study demonstrates that a minimum of 200 states per ReRAM (approximately 8-bit precision) is required to ensure that the degradation in test accuracy is less than 3\%.\\
\indent Two significant sources of non-idealities specific to the thermal neoHebbian synapse include thermal crosstalk and temperature decay. It is noted that test accuracy increases with an increase in $\tau_\mathrm{TH}$ and saturates for $\tau_\mathrm{TH}$ values exceeding 1$\mu s$, as depicted in Fig.\ref{benchmark_simulations}(d). The choice of materials, device dimensions, and crossbar size dictates the $\tau_\mathrm{TH}$ value, and achieving $\tau_\mathrm{TH} \approx 1 \mu s$ is feasible with practical crossbar arrays \cite{srep_3d_crossbar, Yoo2022Tuning2nd}. Thermal crosstalk becomes a critical factor with higher device density, i.e., as minimum feature size ($F$) and crossbar pitch ($K$) decrease. To evaluate the impact of thermal crosstalk, synapse locations are considered from $N\times M$ and $M\times M$ crossbar implementations for the input and recurrent layers, respectively. The data provided in Fig.\ref{numerical_sim}(f,g,h) is used to obtain the thermal coupling coefficient. Fig.\ref{benchmark_simulations}(e) demonstrates that despite notable scaling in $F$ and $K$, the reduction in test accuracy is approximately 3\%. 
Both transistor scaling and thermal crosstalk are critical in determining the scaling potential of thermal synapses. Lastly, Fig.\ref{benchmark_simulations}(f) shows the test accuracy's dependence on memristor variability, showing a decrease of around 1\% for variations up to 100\%. It is shown that increasing the network size results in an improvement in test accuracy. Thus, we attribute the network resilience to various non-idealities to the inherent redundancy in the baseline network \cite{Bellec2019AST} and the implementation of hardware-aware training techniques \cite{IEDM_shu2023}. 
Details on the memristor variability model, its impact on network performance, and the effects of increased ambient temperature on test accuracy are provided in Appendix Section \ref{benchmark_sim}.\\ 
\indent Table \ref{benchmark} compares key metrics of the proposed synapse against the prior works. Per synapse area is determined assuming 1T-1R unit cell configuration, where the heater element is integrated above the ReRAM, resulting in no additional area overhead. The ReRAM cross-sectional area is assumed to be 250$\times$250nm$^2$, with 200 nm spacing between metal lines, giving an estimated cell area of 450F$^2$. Based on 65nm technology for the access transistors, the total cell area is 1.9$\mu$m$^2$. We note that the choice of 65nm technology for access transistor is driven by our fabricated ReRAM’s switching voltages, switching currents, and conductance range \cite{Kim20214KmemristorAP}. However, further reductions unit cell area are possible by decreasing the ReRAM cell area, switching voltages and currents \cite{8776570}. 
The total energy (inference + learning) of the proposed synapse is estimated based on a 10ns spike integration time and a write voltage ($V_\mathrm{w}$) of 1.7V for 1V write pulses. This projection, derived from experimental data, achieves a write pulse duration ($t_\mathrm{PW}$) of 10 ns, resulting in an estimated write energy ($\approx$ $V_\mathrm{w}^{2} G_\mathrm{w} t_\mathrm{PW}$) of $\sim$ pJ, for ReRAM devices. These estimates are supported by previous studies \cite{Kim20214KmemristorAP,6193402}. In conclusion, the proposed synapse offers competitive benefits in terms of both area and energy efficiency.

\definecolor{rowcolor1}{rgb}{1.0, 1.0, 1.0} 
\definecolor{rowcolor2}{rgb}{0.95, 0.95, 0.95} 
\begin{table}[t!]
\centering
\renewcommand{\arraystretch}{1.4} 
\resizebox{\textwidth}{!}{
\begin{tabular}{
    |>{\centering\arraybackslash}m{2.8cm}|
    >{\centering\arraybackslash}m{3cm}|
    >{\centering\arraybackslash}m{2.7cm}|
    >{\centering\arraybackslash}m{2.2cm}|
    >{\centering\arraybackslash}m{2.2cm}|
    >{\centering\arraybackslash}m{2cm}| 
    >{\centering\arraybackslash}m{1.5cm}|}
\hline
     & \textbf{\begin{tabular}[c]{@{}c@{}}Coupling weights\end{tabular}} & \textbf{Eligibility} & \textbf{Per synapse area} & \textbf{Energy per timestep} & \textbf{Eligibility decay time constant} & \textbf{Maturity} \\ \hline
 Y. Demirağ et al.\cite{PCM_eprop}    & PCM conductance                                                      & PCM drift            & $12 \times 12 \, \mu\mathrm{m}^2$             & $\sim 1 \, \mathrm{nJ}^{!}$                            & $\sim \mathrm{s}^{\otimes}$ & +                 \\ \hline
 C. Frenkel et al.\cite{Charlotte_ISSCC_2022}    & CMOS                                                                 & CMOS                 & $1200F^{2 \, \$}$ ($0.94 \, \mu\mathrm{m}^2$)           &   $1.5-178 \, \mathrm{nJ}^{\&}$                             & $\sim \mathrm{ms}^{\boxtimes}$ & ++                \\ \hline
 T. Bohnstingl et al.\cite{Bohnstingl_AICAD_2022} & PCM conductance                                                     & CMOS                 & $5221F^{2 \, \Lambda}$ ($1.02 \, \mu\mathrm{m}^2$)             &   $29 \, \mathrm{nJ}^{\ast}$                         & $\sim \mathrm{ms}^{\boxtimes}$ & +                 \\ \hline
 S. G. Sarwat et al.\cite{Sarwat2021ChalcogenideOF}           & PCM conductance                                                    & Optical response of PCM    & $5 \times 4.8 \, \mu\mathrm{m}^2 \, \Upsilon$
 
 & $\sim \mathrm{mJ}^{\#}$                          & $\sim \mathrm{ns{-}ms}^{\triangle}$ & -                \\ \hline
 \textcolor{blue}{This work}           & \textcolor{blue}{ReRAM conductance}                                                        & \textcolor{blue}{ReRAM local temperature}    & \textcolor{blue}{$450F^2$ ($1.9 \, \mu\mathrm{m}^2$)}            & \textcolor{blue}{$5 \, \mathrm{pJ}$}                    & \textcolor{blue}{$\sim \mathrm{ns{-}\mu s}^{\Omega}$} & \textcolor{blue}{-}                        \\ \hline
\end{tabular}
}
\caption{Comparison of the proposed synapse with prior works. $^{\$}$Calculated assuming SRAM cell area of 150F$^{2}$ in 28nm technology and 8-bit weights. $^{\Lambda}$Calculated assuming 2T1R unit cell and 14nm technology for the access transistor as per \cite{LeGallo2022A6M}. $^\Upsilon$Based on fabricated Ag/GeSe$_{3}$/Ag device dimensions as per \cite{Sarwat2021ChalcogenideOF}. $^{+}$Based on 65nm technology for access transistor and cell area of 450F$^2$. \(^!\)Write energy calculated (\(\approx I^2 t_{pw} / G\)) assuming I\(_{prog}\) \(\sim\) 100\(\mu\)A, \(G \sim\) 1\(\mu\)S, and \(t_{pw} \sim\) 100ns, as per the parameters mentioned in  
\cite{PCM_eprop}. $^\&$Learning energy reported at V= 0.5V \cite{Charlotte_ISSCC_2022}. \(^*\)Only coupling weights are PCM-based; eligibility computations are performed using a high-precision unit. Write energy is estimated roughly according to the values provided in \cite{Bohnstingl_AICAD_2022}: I\(_{prog}\) = 700\(\mu\)A, \(t_{pw}\) = 600ns, \(G = 10\mu\)S. $^\#$Learning energy dominated by optical power \cite{Sarwat2021ChalcogenideOF}. $^\otimes$ Limited by PCM device resistance drift rate. $^\boxtimes$Limited by Von-neumann style sequential computing. $^\triangle$Limited by Ag conductive filament relaxation dynamics. $^\Omega$Limited by thermal time constant}
\label{benchmark}
\end{table}

\section{Discussion \& Summary}
\label{discussion}
\indent The proposed thermal neoHebbian synapse leverages both thermal and electrical effects in computation, forming a multi-physics computing unit. This approach offers several advantages over conventional computation methods. For example, conventional computing units are burdened with converting all signals into the electrical domain, including voltages, currents, and conductances, neglecting other forms of information generated during network operation. By harnessing both electrical and thermal effects in computation, we can maximize the utilization of information derived from network activity, potentially leading to significant improvements \cite{Patel2024HeatassistedNC}.\par 
This approach has gained increasing attention in recent years \cite{computing_with_heat, Kumar2020ThirdorderNE, temp_multi_state_mem, Kim2015Experimental2nd, Yoo2022Tuning2nd, spatio_temporal_correlation}. For instance, Kim et al. \cite{Kim2015Experimental2nd} experimentally demonstrated that the dynamic evolution of internal state variables, particularly temperature, enables ReRAM to mimic Ca$^{2+}$-like dynamics, facilitating the native encoding of temporal information and synaptic weight regulation. They showed that these internal dynamics can be exploited to implement spike-timing-dependent plasticity (STDP) using simple, non-overlapping pulses. Building on this, Yoo et al. \cite{Yoo2022Tuning2nd} proposed material and structural modifications to enhance internal temperature dynamics, validating the concept through the application of STDP-trained spiking neural networks for temporal correlation detection tasks. Another study \cite{spatio_temporal_correlation} explored the use of thermal crosstalk in neuromorphic computing, proposing its potential for future applications. In related work, it shows that the thermal crosstalk-driven spatiotemporal communication in multiple Mott neurons achieves energy efficiency several orders of magnitude greater than state-of-the-art digital processors \cite{computing_with_heat}. Similarly, Kumar et al. \cite{Kumar2020ThirdorderNE} leveraged thermal dynamics to demonstrate 15 distinct neuronal behaviors using nanoscale third-order circuit elements, showing promise for the development of highly efficient neuromorphic hardware.\par
While multi-physics computing units, particularly those involving temperature, offer significant advantages, they also present several practical challenges. Unlike measurable variables such as current or voltage, temperature is a hidden variable, making direct measurement and control difficult. Furthermore, elevated temperatures can accelerate device degradation and lead to various reliability issues \cite{Temp_reliability,thermal_managment}. Although this work and several other works \cite{computing_with_heat, Kumar2020ThirdorderNE, temp_multi_state_mem, Kim2015Experimental2nd, Yoo2022Tuning2nd, spatio_temporal_correlation} demonstrated a method of exploiting thermal effects for computation, significant challenges remain for future real-world applications. \par 
In addition, heat dissipation is an unavoidable byproduct of electronic system operation, and it is increasingly pronounced as devices continue to shrink in size \cite{Salahuddin2018TheEO}. While efforts to reduce power dissipation remain a priority, exploring innovative approaches to harness electro-thermal effects could unlock new possibilities. For example, such approaches could drive advancements in novel materials with tailored thermal properties, where electronic and thermal behaviors can be independently controlled. Moreover, the development of nanoscale devices capable of regulating heat flow, such as thermal diodes or thermal transistors, presents promising directions for future research \cite{Yang2023SolidStateET, Wei2022ElectriccontrolledTT}.\par
In summary, we have proposed and experimentally validated ReRAM-based neo-Hebbian synapses. The performance improvements provided by these synapses were evaluated through two representative applications based on the scalable e-prop learning algorithm. Our findings demonstrate that the proposed thermal neo-Hebbian synapses significantly reduce both time-to-solution and energy-to-solution. This underscores their potential for facilitating fast, scalable, online, and robust learning in neuromorphic hardware.
\section*{Acknowledgements} 
The authors would like to express their gratitude for the valuable discussions with R. Legenstein and H. Kim. This work has received support from ONR grant \#N00014-22-1-2842 and the Bekker programs. Additionally, S. Pande acknowledges financial support from the Fulbright-Nehru Doctoral Research Fellowship and the Government of India through the Prime Minister Research Fellowship.

\section*{Conflict of Interest}
The authors declare no conﬂict of interest.

\section*{Data Availability Statement}
The data that support the ﬁndings of this study are available from the corresponding author upon reasonable request.
\medskip
\medskip
\bibliographystyle{msp}
\bibliography{ref}
\section{Appendix}

\subsection{Prior Works}
\label{prior_works}
One notable non-ideality in phase-change memory (PCM), namely resistance drift, has been leveraged for eligibility computation \cite{PCM_eprop}. Specifically, two PCM devices were employed within each synapse: one to implement the eligibility trace by utilizing the inherent resistance drift behavior of PCM, and the other to encode the coupling weight through its non-volatile conductance \cite{PCM_eprop}. However, this approach faces common challenges associated with PCM-based synapses, including significant day-scale drift toward the amorphous state, even in devices optimized for high retention, and an excessively abrupt reset transition. To address these issues, a differential pair synapse with a set-only programming scheme was proposed. However, this design requires at least eight CMOS transistors per synapse, resulting in a large footprint and sparse synapse integration. In a separate study, the combined influence of electrical and optical stimuli on the switching behavior of non-volatile Ag/GeSe$_{3}$/Ag memristive devices was explored \cite{Sarwat2021ChalcogenideOF}. For example, when an electrical pulse was applied to an illuminated device, it spontaneously transitioned to a conducting state. In contrast, no resistive switching occurred if the device was illuminated without a voltage stimulus, or if a sub-threshold electrical stimulus was applied in dark conditions. Utilizing this behavior, the authors demonstrated the implementation of a three-factor learning rule. While these devices hold promise for high-density integration, energy consumption could be a limiting factor, with total energy consumption primarily driven by optical power, typically within the milliwatt range \cite{Sarwat2021ChalcogenideOF}. 
Other research has focused on using PCM devices solely for storing coupling weights, employing high-precision computational units for eligibility trace computation \cite{Bohnstingl_AICAD_2022, ortner2024learning}, or utilizing mixed-signal analog/digital neuromorphic circuits \cite{Charlotte_ISSCC_2022}. However, both approaches face challenges related to transistor scaling, especially for large-scale network implementations. Consequently, there remains significant scope for developing neo-Hebbian synapses that harness memristor physical dynamics for the hardware implementation of advanced learning rules.

\subsection{e-prop Key Equations}
\label{e_prop_eq}
The LIF neuron's membrane potential and spike generation are characterized by Eq.(\ref{eq1}-\ref{eq3}). Specifically, LIF neurons' membrane potential ($V^{'}_{j}(t)$) is updated at fixed discrete time step $t$ following the Eq.(\ref{eq1}) \& Eq.(\ref{eq2}). The exponential decay factor in the first term of Eq.(\ref{eq1}) describes the behavior of the membrane potential in the absence of spikes, and the last two terms denote the contribution from recurrent weights ($w^{h}$) and input weight ($w^{i}$), respectively. Here, $\tau_{m}$ denotes the neuron membrane potential time constant. Whenever $V^{'}_{j}(t)$ surpasses the threshold voltage (V$_\mathrm{th}$), it is reduced by an amount equal to the threshold voltage, and an output spike $s(t)$ is generated, as described in Eq.(\ref{eq3}). Otherwise, $V^{'}_{j}(t)$ follows the behavior outlined in Eq.(\ref{eq1}), and no output spike is generated.
\begin{equation}
V_j^{\prime}(t) = \exp \left[-\frac{1}{\tau_m}\right] V_j(t-1) + \sum_{i, i \neq j}^N w_{i j}^h(t) s_i^h(t) - w_{j j}^h(t) s_j^h(t) + \sum_i^M w_{i j}^{i}(t) s_i^i(t)
\label{eq1}
\end{equation}

\begin{minipage}{0.5\textwidth}
\begin{equation}
V_j(t) = \left\{
\begin{array}{cl}
V_j^{\prime}(t)-V_\mathrm{th}, & \text{if } V_j^{\prime}(t) \geq V_\mathrm{th}\\
V_j^{\prime}(t), & \text{otherwise}
\end{array}
\right.
\label{eq2}
\end{equation}
\end{minipage}%
\begin{minipage}{0.5\textwidth}   
\begin{equation}
s_j(t+1) = \left\{
\begin{array}{cc}
1, & \text{if } V_j^{\prime}(t) \geq V_\mathrm{th}\\ 
0, & \text{otherwise}
\end{array}\right.
\label{eq3}
\end{equation}
\end{minipage} 
\vspace{0.5cm}

\noindent Network loss (L$_j$) is determined at output neuron $k$ by assessing how much the output $y_k$ differs from the target value $y^{target}_k$ at time $t$ following Eq.(\ref{eq4}). Batch-mode stochastic gradient descent is used to update the output layer weights ($w^o$) following Eq.(\ref{eq5}). 

\begin{minipage}{0.5\textwidth}
\begin{equation}
L_{\mathrm{j}}(\mathrm{t}) = \sum_{\mathrm{k}=1}^{\mathrm{C}} \mathrm{w}_{\mathrm{jk}}^{\mathrm{o}}\left(\mathrm{y}_{\mathrm{k}}(\mathrm{t})-\mathrm{y}_{\mathrm{k}}^{\mathrm{target}}(\mathrm{t})\right) 
\label{eq4}
\end{equation}
\end{minipage}%
\begin{minipage}{0.5\textwidth}   
\begin{equation}
\Delta w_{j k}^o = \eta \sum_{t=1}^U s_j^h(t)\left(y_k(t)-y_k^{\text {target }}(t)\right)
\label{eq5}
\end{equation}
\end{minipage}
\vspace{0.5cm}

\noindent Eligibility state ($f(t)$) and pseudo-gradient ($\psi(t)$) are modeled using Eq.(\ref{eq6}) and Eq.(\ref{eq7}), respectively. The function $f(.)$ denotes low pass filtering action. Eq.(\ref{eq8}) represents the accumulated eligibility value (e$_{\sum}$) over U-steps, i.e., one dataframe presentation cycle. Neohebbian synapse weights are updated at the end of the dataframe according to Eq.(\ref{eq9}). $\eta$ represents learning rate.

\begin{minipage}{0.5\textwidth}
\begin{equation}
f(t) = \exp \left[-\frac{1}{\tau_m}\right] f(t-1) + s(t)
\label{eq6}
\end{equation}
\end{minipage}%
\begin{minipage}{0.5\textwidth}   
\begin{equation}
\psi(t) = \frac{\beta}{V_\mathrm{th}} \max \left[0,1-\left|V(t-1) / V_\mathrm{th}-1\right|\right] L(t)
\label{eq7}
\end{equation}
\end{minipage}

\begin{minipage}{0.5\textwidth}
\begin{equation}
e_{\Sigma} = \sum_{t=1}^U e(t), \text{ where } e(t) = f(t) \times \psi(t)
\label{eq8}
\end{equation}
\end{minipage}%
\begin{minipage}{0.5\textwidth}   
\begin{equation}
\Delta w^{i / h} = \eta e_{\Sigma}
\label{eq9}
\end{equation}
\end{minipage}

\noindent The initial values for $V^{'}_{j}(t)$, $f(t)$, $\psi(t)$,  $e_{\sum}$ are all set to zero at $t$ =0, i.e., before the presentation of the new training dataframe. Updates to the readout weights and input/recurrent neoHebbian weights occur exclusively at the end of each dataframe.  The values of hyperparameters used  are V$_\mathrm{th}$ = 0.615, $\tau_{m}$ = 200ms, learning rates $\eta$ = 0.1 $\div$ 1, $\beta$ = 0.3. A discrete time step of $\delta$t = 1ms is used for all simulations.

\subsection{Eligibility Computation at Synapse}
\label{elig_comp}
\begin{figure}[b!]
\centering
\hspace{-0.25cm}
\includegraphics[trim={0 0 0 0}, clip ,width=0.4\textwidth]{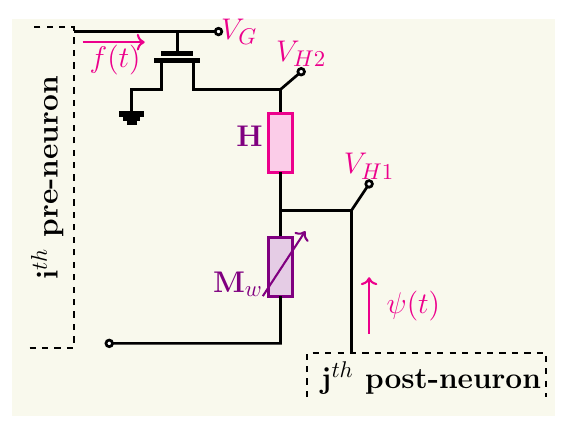} 
\caption{Schematic of the 1$T$-1$H$-1$M$ unit cell showing biasing conditions during the e-update phase.}
\label{e_update_appendix}
\end{figure}
\noindent In the following, we refer to the 1$T$-1$H$-1$M$ synaptic cell and the specific notation, as shown in Fig.\ref{e_update_appendix}. In the e-update phase, the transistor operates in the triode region. Thus, the current flowing through the series combination of the transistor and heater is given by eq.(\ref{eq10}). 
\begin{equation}
    I_\mathrm{D}(t)=\mu_{eff} C_\mathrm{eff} \frac{W}{L}\left(V_\mathrm{GS}(t)-V_\mathrm{TH}\right) V_\mathrm{H2}(t)
    \label{eq10}
\end{equation}
We express $V_\mathrm{H2}(t)$ in terms of $V_\mathrm{H1}(t)$, the heater's electrical resistance ($R$), and transistor-related parameters such as carrier mobility ($\mu_{eff}$), gate capacitance ($C_\mathrm{eff}$), threshold voltage ($V_\mathrm{TH}$), transistor width ($W$), and channel length ($L$) in eq.(\ref{eq11}),
\begin{equation}
    V_\mathrm{H2}(t)=\frac{V_\mathrm{H1}(t)}{1+R k V_\mathrm{OV}(t)} 
    \label{eq11}
\end{equation}

\begin{equation}
V_\mathrm{OV}=V_\mathrm{GS}-V_\mathrm{TH} \quad \& \quad   k=\mu_{eff} C_\mathrm{eff} \frac{W}{L}
\end{equation}

\noindent Following that, the power dissipated across the heater can be expressed as in eq.(\ref{eq12}). 
\begin{equation}
P_\mathrm{H}=\frac{1}{R} V_\mathrm{H1}^2 \frac{\left(R k V_\mathrm{OV}\right)^2}{\left(1+R k V_\mathrm{OV}\right)^2}
\label{eq12}
\end{equation}
Now, the voltage signals $V_\mathrm{G}$ and $V_\mathrm{H1}$ are scaled as shown in eq.(\ref{eq13}) to ensure that $P_\mathrm{H}$ is proportional to the product of $f(t)$ and $\psi(t)$. 
\begin{equation}
V_\mathrm{H1} \propto \sqrt{\psi(t)} \quad \& \quad V_\mathrm{OV} \propto \frac{\sqrt{f(t)}}{\operatorname{\textit{Rk}}\;(1-\sqrt{f(t)})}  
\label{eq13} 
\end{equation}
Let $T^0(t)$ represent the increase in heater temperature above ambient temperature ($T_\mathrm{amb}$) at time $t$, as defined in eq.(\ref{eq14}). The temperature change $\Delta T^0(t)$ is proportional to the power dissipated in the heater, $P_\mathrm{H}$, and therefore also proportional to the product of $f(t)$ and $\psi(t)$. Due to thermal coupling, the local temperature of $M_\mathrm{w}$ follows the same proportionality, as shown in eq.(\ref{eq15}). In this context, $t_\mathrm{PW}$, $\tau_\mathrm{TH}$, and $R_\mathrm{TH}$ refer to the pulse width, thermal time constant, and thermal resistance, respectively.

\begin{equation*}
T^0(t)=T(t)-T_\mathrm{amb}
\end{equation*}

\begin{equation}
T^0(t)=T^0(t-1)+\frac{t_\mathrm{PW}}{\tau_\mathrm{TH}}\left[P_H(t) R_\mathrm{T H}-T^0(t-1)\right] 
\label{eq14}
\end{equation}

\begin{align*}
\begin{split}
T^0(t)-T^0(t-1) = \Delta T^0(t) = \frac{t_\mathrm{PW}}{\tau_\mathrm{TH}}[P_\mathrm{H}(t) R_\mathrm{T H} - \;T^0(t-1)]
\end{split}
\end{align*}

\begin{equation}
    \Delta T^0(t) \propto P_\mathrm{H}(t) \propto f(t) \times \psi(t)
    \label{eq15}
\end{equation}

\begin{figure}[b!]
\centering
\includegraphics[trim={0cm 0cm 0cm 0cm}, clip ,width=0.7\textwidth]{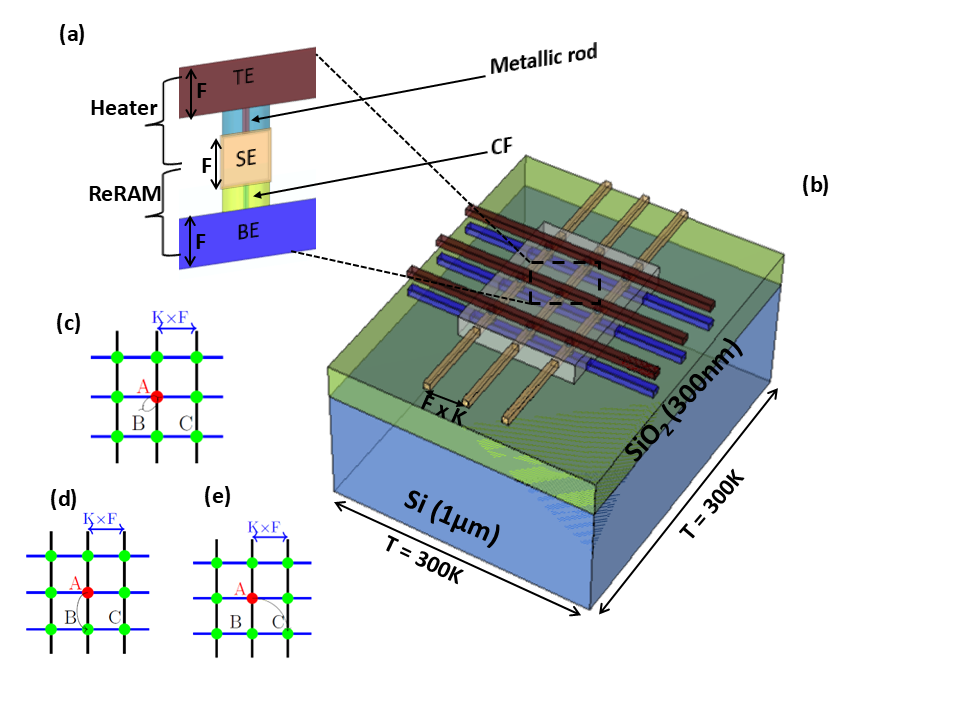} 
\caption{(a) Cross-sectional schematic of the modeled heater-integrated ReRAM structure. (b) The modeled geometry is incorporated into a 3×3 crossbar array, where F denotes the minimum feature size, and K denotes the crossbar pitch. Oxide thickness for the heater and ReRAM is t$_{ox}$ = 30nm. (c) A scenario where a heating pulse is applied to the heater at position A and thermal coupling is measured for the ReRAM at position A. Similarly, (d, e) shows the thermal coupling between the heater at position A and the ReRAM at positions B and C, respectively.}
\label{crosbar_numerical}
\end{figure}

\subsection{Numerical Modeling}
\label{num_model}
Fig.\ref{crosbar_numerical}(a) illustrates the schematic of the heater-integrated ReRAM device, which is incorporated into a crossbar array, as shown in Fig. \ref{crosbar_numerical}(b). The temperature distribution within the crossbar array is obtained by solving the transient heat flow eq.(\ref{eq16}), while the potential distribution is determined using eq.(\ref{eq17}). Both equations are solved self-consistently across the entire simulation domain using a COMSOL multi-physics solver.

\begin{equation}
\rho\;C\;\frac{\partial T}{\partial t} = \nabla\kappa_\mathrm{th}\;\nabla T + \sigma |\nabla E|^2    
\label{eq16}
\end{equation}

\begin{equation}
E = -\nabla V 
\label{eq17}
\end{equation}
The parameters $C$, $\rho$, $\kappa_\mathrm{th}$, and $\sigma$ denote the material's specific heat capacity, density, thermal conductivity, and electrical resistivity, respectively. Voltage and electric field are indicated by $V$ and $E$, respectively. Parameter values used in this study are shown in the table \ref{table_3}.  

To access the heater and ReRAM devices, a voltage is applied to one of the vertical planes of the respective electrodes, while all other electrode regions remain electrically insulating. This is achieved by imposing a Neumann boundary condition, $n.j=0$, ensuring no current flows through these insulating boundaries. The thermal boundary conditions assume that the bottom side of the substrate functions as an ideal heat sink, maintained at a constant temperature of 300 K, whereas all other surfaces are considered thermally adiabatic.
The metallic nanorod acting as a heater is assumed to have a radius of 4.19nm, while the ReRAM conductive filament (CF) has a radius of 3.1nm. To evaluate the thermal coupling between the heater and the ReRAM, a heating pulse is applied to the heater, and the resultant passive temperature rise in the neighboring ReRAM is measured. The thermal coupling coefficient is defined as the ratio of the temperature rise at the synapse (ReRAM) to the temperature rise at the heater in response to the applied heating pulse. For instance, as depicted in Fig.\ref{crosbar_numerical}(c), the thermal coupling between the heater and ReRAM positioned at intersection A is illustrated. Similarly, Fig.\ref{crosbar_numerical}(d) and Fig.\ref{crosbar_numerical}(e) denote the thermal coupling between the heater at intersection A and ReRAM positioned at intersections B and C, respectively. It is desired to maximize thermal coupling in the scenario presented in Fig.\ref{crosbar_numerical}(c), while minimizing it in the scenarios illustrated in Fig.\ref{crosbar_numerical}(d) and Fig.\ref{crosbar_numerical}(e). The modified structure presented in Fig.\ref{numerical_sim}(e) aims to achieve the same. Specifically, it reduces the distance between the heater and ReRAM to enhance desired self-thermal coupling. It increases the thickness of both the top and bottom electrodes to slow the propagation of heat flux toward neighboring ReRAM, thereby reducing unintentional thermal crosstalk. Except for these geometrical changes, all electrical and thermal boundary conditions, as well as material parameters, remain the same for the modified structure. Finally, Fig.\ref{numerical_sim}(f,g,h) compares the thermal coupling coefficients for baseline structure (shown in Fig.\ref{numerical_sim}(b)) and modified structure (shown in Fig.\ref{numerical_sim}(e)).
\definecolor{rowcolor1}{rgb}{0.9, 0.9, 0.9} 
\definecolor{rowcolor2}{rgb}{1.0, 1.0, 1.0} 

\begin{table}[h!]
\centering
\setlength{\extrarowheight}{5pt} 
\resizebox{0.8\textwidth}{!}{
\begin{tabular}{
    |>{\centering\arraybackslash}m{2.5cm}|
    >{\centering\arraybackslash}m{2.5cm}|
    >{\centering\arraybackslash}m{2.5cm}|
    >{\centering\arraybackslash}m{2.5cm}|
    >{\centering\arraybackslash}m{2.5cm}|
    >{\centering\arraybackslash}m{2.5cm}|}
\hline
\textbf{Parameters} & \textbf{Swicthing oxide} & \textbf{Metallic rod} & \textbf{Electrode} & \textbf{CF} & \textbf{Isolation oxide} \\ \hline 
$\kappa_\mathrm{th}$ [W/mK] & 1.4 & 23 & 71.8 & 23 & 0.5 \\ \hline 
$\sigma$ [S/m] & 1e-6 & 1e5 & 1e7 & 1e5 & 1e-6 \\ \hline 
$\rho$ [kg/m$^3$] & 6850 & 5200 & 12033 & 6850 & 745 \\ \hline 
C [J/kg K] & 306 & 450 & 244 & 306 & 2200 \\ \hline
\end{tabular}
}
\caption{Parameter values used in the numerical simulations. For the purpose of simulation, the metallic rod and CF are assumed to have the same electrical and thermal properties. The material parameters used in the simulations are not intended to restrict the implementation of the proposed synapse to specific material parameter values. Instead, they are representative and used to study the scope/limitations of the proposed synapse. These values are in agreement with values used in prior works \cite{PANDE2023108636,caliberation,caliberation2,we_lu_acs_nano,multi-physics}. }
\label{table_3}
\end{table}

\begin{figure}[t!]
\centering
\hspace{-0.25cm}
\includegraphics[trim={2cm 0cm 2cm 0cm}, clip ,width=0.6\textwidth]{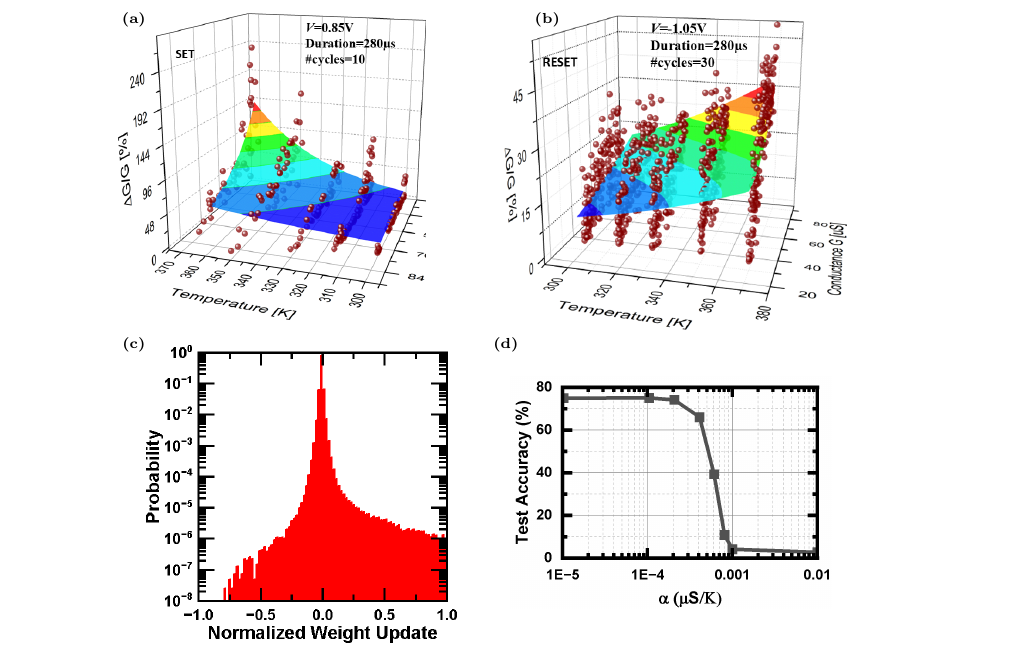} 
\caption{Experimental data overlaid with modeled $\Delta G$ as a function of temperature and $G_{0}$ for (a) SET (b) RESET process. (c) Histogram showing the probability of normalized weight update during training. (d) Test accuracy comparison against various values of $\alpha$.}
\label{hist_gram}
\end{figure}

\subsection{Benchmark simulations}
\label{benchmark_sim}
The experimental data presented in Fig.\ref{measured_data}(b) and Fig.\ref{measured_data}(c) is modeled using Eq.\ref{fit_func1} and used in hardware-aware network simulations.
\begin{equation}
    \frac{\Delta G}{G_{0}} = (a\;e^{b\;G_{0}})\;T^{(c\;e^{d\;G_{0}})}.
\label{fit_func1}
\end{equation}
Here, $T$ represents the ambient temperature. Fig.\ref{hist_gram}(a) and Fig.\ref{hist_gram}(b) show the color-mapped surface obtained using Eq.\ref{fit_func1}, overlaid on the measurement data. The fitting parameters $a$, $b$, $c$, and $d$ for the SET (RESET) process are 0.143 (0.3124), 2.216 (0.8064), 0.8232 (1.138), and 0.4043 (-0.8806), respectively.

\subsubsection{Case Study\#2: Benchmark Simulations}
\label{temp_cond}
\noindent 
To account for device-device variations, the Gaussian noise is added to fitting parameters $a$,$b$,$c$,$d$, 
\begin{equation*}
    [a,b,c,d] = \mathcal{N}([a,b,c,d], D_{v}).
\end{equation*}
Here, $D_{v}$ accounts for device-to-device variability, and the mean values of [$a$,$b$,$c$, $d$] are obtained by fitting the experimental data using Eq.(\ref{fit_func1}). Subsequently, to account for cycle-to-cycle variations, we add the Gaussian noise to the calculated $\Delta G$ value as follows,
\begin{equation*}
    \Delta G = \mathcal{N}(\Delta G, C_v\;\Delta G).
\end{equation*}
The parameter \(C_v\) represents cycle-to-cycle variability, which in our model scales with \(\Delta G\). This variability becomes more pronounced at higher \(\Delta G\) values.\\ 

\noindent We plot a histogram showing the probability of \(\frac{\Delta G}{G_{0}}\) update during training to probe the impact of variability on test accuracy. Fig.\ref{hist_gram}(c) shows a histogram with the \(\frac{\Delta G}{G_{0}}\) on the x-axis and the probability of weight update on the y-axis. Our simulations indicate that smaller \(\Delta G\) values occurrence is more likely. This means that conductance tends to change by small amounts more frequently during training. This behavior mitigates the impact of cycle-to-cycle variability. \\

\noindent The rise in local ReRAM temperature can affect its programmed conductance, which we model using \( W = W(1 + \alpha(T - T_\mathrm{amb})) \), where \( W \) represents the ReRAM conductance, and \( \alpha \) is the temperature coefficient \cite{9047174}. Our observations indicate that test accuracy drops sharply when \( \alpha \) exceeds \( 4 \times 10^{-4} \, \mu\text{S}/\text{K} \). While smaller \( \alpha \) values are ideal, they are influenced by both material and geometric factors. Therefore, selecting appropriate device materials and geometrical parameters is crucial \cite{9047174,temp_multi_state_mem}.


\end{document}